\documentclass[aps,pra,reprint,superscriptaddress]{revtex4-1}
\usepackage{graphicx}
\usepackage{amsmath}
\usepackage{amssymb}
\usepackage{color}
\usepackage{mathrsfs}
\usepackage{dsfont}
\usepackage[defaultlines=3,all]{nowidow}

\newcommand*{\p}{\partial}

\renewcommand*{\le}{\left}
\newcommand*{\ri}{\right}
\newcommand*{\twovector}[2]{\left(\begin{array}{c}
    #1\\
    #2
  \end{array}\right)}
\newcommand*{\twomatrix}[4]{\left[\begin{array}{cc}
    #1 & #2\\
    #3 & #4
  \end{array}\right]}

\begin{document}
\graphicspath{{./figures/}}
\title{Phase-tunable Majorana bound states in a topological N-SNS junction}

\author{Esben Bork Hansen}
\affiliation{Center for Quantum Devices, Niels Bohr Institute, University of Copenhagen, DK-2100 Copenhagen \O, Denmark}
\author{Jeroen Danon}
\affiliation{Center for Quantum Devices, Niels Bohr Institute, University of Copenhagen, DK-2100 Copenhagen \O, Denmark}
\affiliation{Niels Bohr International Academy, Niels Bohr Institute, University of Copenhagen, DK-2100 Copenhagen \O, Denmark}
\author{Karsten Flensberg}
\affiliation{Center for Quantum Devices, Niels Bohr Institute, University of Copenhagen, DK-2100 Copenhagen \O, Denmark}

\date{\today}
\begin{abstract}
We theoretically study the differential conductance of a one-dimensional normal-superconductor-normal-superconductor (N-SNS) junction with a phase bias applied between the two superconductors. We consider specifically a junction formed by a spin-orbit coupled semiconducting nanowire with regions of the nanowire having superconducting pairing induced by a bulk $s$-wave superconductor. When the nanowire is tuned into a topologically non-trivial phase by a Zeeman field, it hosts zero-energy Majorana modes at its ends as well as at the interface between the two superconductors. The phase-dependent splitting of the Majorana modes gives rise to features in the differential conductance that offer a clear distinction between the topologically trivial and non-trivial phases. We calculate the transport properties of the junction numerically and also present a simple analytical model that captures the main properties of the predicted tunneling spectroscopy.
\end{abstract}

\maketitle

\section{\label{sec:intro}Introduction}

The edges of one- and two-dimensional spinless or spin-triplet $p$-wave superconductors can host localized zero-energy excitations, so-called Majorana bound states (MBSs) or Majorana zero-modes---named such because their second quantized operators are self-adjoint~\cite{JETPL.70.609,PhysRevB.61.10267,1063-7869-44-10S-S29}. These MBSs obey non-Abelian exchange statistics, and the notion that they could therefore be utilized for topologically protected quantum computation~\cite{RevModPhys.80.1083,PhysRevLett.86.268} sparked an intense search for feasible realizations of the $p$-wave topological superconducting phase. Concrete proposals included using superfluid states of cold fermionic atoms \cite{PhysRevLett.94.230403,PhysRevLett.98.010506,PhysRevLett.103.020401}, heterostructures consisting of topological insulators, $s$-wave superconductors, and ferromagnetic insulators \cite{PhysRevLett.100.096407,PhysRevLett.103.107002}, or non-centrosymmetric superconductors with strong (Rashba) spin-orbit coupling~\cite{Lee2009,PhysRevB.79.094504}.

Arguably, one of the simplest and most promising ideas how to create an effective $p$-wave superconductor is to use a quasi-one-dimensional semiconducting nanowire with proximity-induced $s$-wave superconductivity and strong spin-orbit coupling, subjected to a magnetic field perpendicular to the effective spin-orbit field~\cite{PhysRevLett.105.177002,PhysRevLett.105.077001}. As the magnetic field increases, the induced gap in the wire is expected to close and reopen, which signals a transition from a topologically trivial phase to a non-trivial (effectively $p$-wave superconducting) phase, where localized zero-energy MBSs form at the ends of the wire. These proposals were rapidly followed by pioneering experiments, where tunneling spectroscopy into one end of the wire indeed showed a zero-bias peak appearing when the magnetic field was increased~\cite{Mourik25052012, Das2012, PhysRevB.87.241401}, which was interpreted as a signature of the formation of MBSs.

However, several of the experimental observations seemed to be incompatible with the interpretation in terms of a topological phase transition. For instance, a clear closing and reopening of the gap was never observed in experiment, and also the measured height of the zero-bias conductance peak was only a fraction of the theoretically predicted value of $2e^2/h$. Soon it was pointed out that these ``inconsistencies'' could be attributed to realistic complications such as inhomogeneous depletion profiles, population of more than one transverse subband, disorder, or finite temperature~\cite{PhysRevB.86.180503,Rainis2013}; and within all these explanations the zero-bias peak could still be associated with MBSs living at the ends of the wire. But also interpretations of the data {\it without} zero-energy MBSs were proposed: Robust zero-bias peaks in the conductance of a topologically trivial wire can, for instance, appear due to an interplay of disorder, finite temperature, and population of multiple subbands~\cite{PhysRevLett.109.267002}, due to smooth confining potentials~\cite{PhysRevB.86.100503}, or due to the Kondo effect~\cite{PhysRevLett.109.186802}.

The origin of the observed zero-bias peaks is thus still under debate, and identifying other, more distinguishing signatures of MBSs in semiconducting nanowires would therefore be very useful. So far, proposals include using a superconductor-normal-superconductor (SNS) setup to focus on the $4\pi$-periodic Josephson current~\cite{1063-7869-44-10S-S29,ramon1} or the critical Josephson current through a multiband wire~\cite{ramon2}, measuring the actual localization of the zero-energy states at the ends of the wire, and observing the characteristic oscillations of the splitting of the zero-bias peak as a function of increasing magnetic field~\cite{PhysRevB.86.220506,PhysRevB.86.180503}. The observation of a $4\pi$-periodic Josephson current can be hampered by the finite overlap of the MBSs, leading again to a regular $2\pi$-periodic current~\cite{ramon1}. Probing the localization of the zero-energy mode would require measuring the local density of states in the wire by e.g.\ scanning tunneling microscopy, which cannot readily be performed on these wires. The characteristic oscillations of the split zero-bias peak have also proven to be difficult to observe: Temperature broadening of the energy levels and a soft induced superconducting gap make it hard to clearly distinguish the oscillations of the MBS's energy in the topologically non-trivial phase from zero-bias crossings that are also present in the topologically trivial phase. The most decisive measurement would of course be the observation of non-Abelian braiding statistics, but even the simplest braiding scheme requires a challenging experimental setup and delicate control~\cite{PhysRevB.88.035121, PhysRevB.84.094505, PhysRevLett.106.090503}. A more tractable experiment that would allow for the distinction between the different possible origins of the zero-bias conductance peak is therefore a preferable stepping stone before moving on to more complicated braiding experiments. 

\begin{figure}
  \includegraphics[scale=0.95]{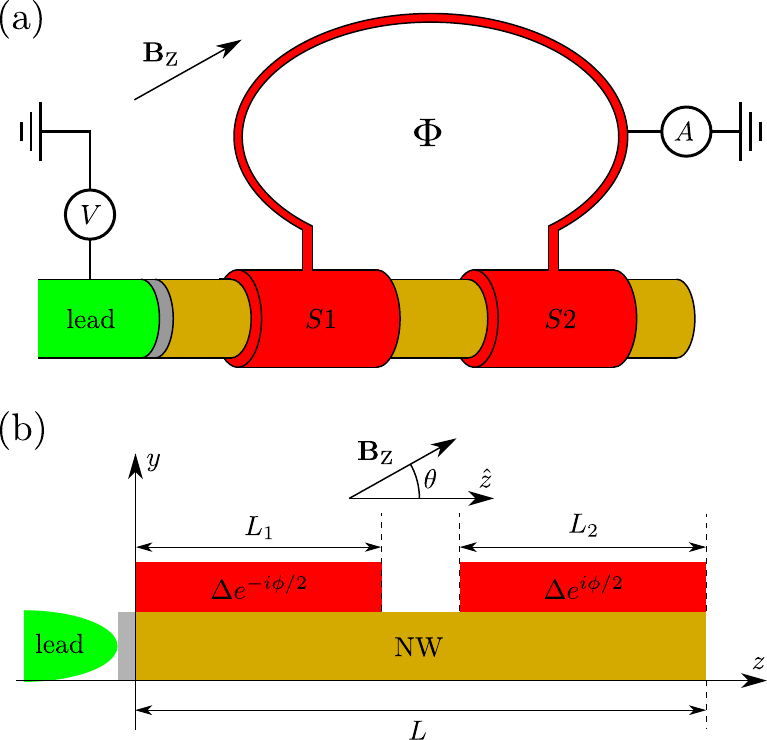}
  \centering
  \caption{(Color online) (a) Cartoon of the N-SNS setup we consider. A semiconducting nanowire with epitaxially grown $s$-wave superconductor covering two separate parts of the wire. The two superconductors $S1$ and $S2$ are connected by a grounded superconducting loop through which a flux $\Phi$ can be threaded, resulting in a phase difference between $S1$ and $S2$. An external magnetic field ${\bf B}_{\rm Z}$ is applied that couples to the electronic spins inside the wire, but does not add to $\Phi$. The wire is tunnel coupled to a probe lead at the left end which can be used for tunneling spectroscopy by applying a bias voltage $V$ to it while measuring the current $A$. (b) More schematic view of the setup. The wire is oriented along the $z$-axis and ${\bf B}_{\rm Z}$ lies in the $yz$-plane, where the effective spin-orbit field is assumed to point along the $y$-axis.\label{fig1} }
\end{figure}
In this work we investigate in more detail the SNS setup, where two separate parts of the wire are coupled to two separate bulk superconductors, interconnected by a superconducting loop to control the phase difference between them (see the sketch in Fig.~\ref{fig1}).
Under the right circumstances there can emerge up to four low-energy MBSs (one at each NS boundary inside the wire), which are known to have very characteristic phase-dependent properties~\cite{ramon1,ramon3}.
Inspired by this, we suggest to study the differential conductance measured by an additional normal tunnel probe coupled to one end of the nanowire, rather than investigating the Josephson current in the junction~\cite{ramon1,ramon2}.
Therefore the superconducting loop is needed to create a controllable phase difference across the junction, and it thus replaces the voltage bias inducing a time-dependent phase difference in Ref.~\cite{ramon1} or the supercurrent bias in Ref.~\cite{ramon2}.
We thus investigate a normal-superconductor-normal-superconductor (N-SNS) junction, and look for signatures of the MBSs in the differential conductance as measured through the probe.
Creating such a topological SNS junction is feasible with the current state of experimental techniques, and probing by means of tunneling spectroscopy is standard practice in this field.
We show that such an experiment should in principle allow to determine whether or not measured low-bias peaks in the differential conductance can be associated with non-local fermionic states formed by MBSs inside the wire, and we believe that at present our proposal constitutes one of the most straightforward ways to experimentally verify the emergence of topological phases in a spin-orbit-coupled semiconducting nanowire.

In Section \ref{sec:model} we describe the setup we have in mind and present a Hamiltonian to model this system. We explain how one can calculate the differential conductance of this system from our model Hamiltonian, and in Section \ref{sec:results} we present our numerical results. We examine the conductance of the wire for energies smaller than the superconducting gap, calculating the conductance spectrum as a function of applied magnetic field and of the phase difference between the two superconductors. We compare the spectra that result when the wire is in a topological phase and in a trivial phase, and we find that the conductance spectrum, when tunneling into one end of the wire, is only sensitive to the phase difference between the superconductors if the wire is in a topological phase. Depending on the phase difference and the applied magnetic field, one to four low-energy subgap conductance peaks are visible: For zero phase difference, one strong zero-bias peak appears after entering the topological phase; at higher magnetic fields this peak splits and starts to oscillate due to the increasing overlap of the two MBSs at the ends of the wire. For a phase difference close to $\pi$ we can see two more low-energy peaks, which we associate with the formation of two additional MBSs close to the central normal region of the N-SNS junction, where the superconducting phase  changes sign. In Section \ref{sec:effective_model} we support our interpretation with a simple low-energy model based on a one-dimensional spinless $p$-wave superconductor with a phase discontinuity. This toy Hamiltonian reproduces all important qualitative features of our numerical results and provides insight in the structure of the ``Majorana subspace'' including the gradual gapping out of the central two Majorana states when the phase difference is reduced to zero. Finally, in Section \ref{sec:conc} we present our conclusions.

\section{Model\label{sec:model}}

We consider a semiconducting nanowire proximity coupled to two $s$-wave superconductors, as illustrated in Fig.~\ref{fig1}(a). The two superconductors are connected by a superconducting loop such that the phase difference $\phi$ between them can be tuned by threading a flux $\Phi$ through the loop.
Further, the wire is assumed to have strong spin-orbit interaction and we include a magnetic field that results in a Zeeman splitting of the electronic states inside the wire but does not add to the flux $\Phi$.
At its left end, the wire is tunnel coupled to a normal-metal probe which can be used to measure the differential conductance of the system, as indicated in Fig.~\ref{fig1}(a).

A more schematic picture of the setup is shown in Fig.~\ref{fig1}(b). The nanowire of length $L$ is oriented along the $z$-axis and it is proximity coupled to two bulk $s$-wave superconductors of length $L_1$ and $L_2$ which have the effective pairing potentials $\Delta e^{-i\phi/2}$ and $\Delta e^{i\phi/2}$ respectively. We assume the effective spin-orbit field to point along the $y$-axis and the Zeeman field to lie in the $yz$-plane, $\mathbf{B}_{\rm Z} = B (\hat{z}\cos\theta + \hat{y}\sin\theta)$. By varying the direction of this field we can investigate both the topologically trivial state and non-trivial state of the system: The wire is in a trivial state when the Zeeman field is parallel to the spin-orbit field ($\theta = \pm \pi/2$), and can be in a topological state for non-parallel fields~\cite{rex}.

To model this system, we use the Hamiltonian
\begin{equation}
H = H_{\text{NW}} + H_1 + H_2  + H_{\text{t}},\label{eq:hw}
\end{equation}
where $H_{\text{NW}}$, $H_1$, and $H_2$ describe the electrons in the nanowire and the two superconductors respectively, and $H_{\text{t}}$ accounts for the coupling between the different parts of the system.

Assuming that the wire is thin enough such that only the lowest electronic subband is occupied, we write a one-dimensional Hamiltonian in a Bogoliubov-de Gennes (BdG) form, $H_{\text{NW}} = \frac{1}{2}\int dz\, \boldsymbol \Psi^\dag(z) {\cal H}_{\rm NW} \boldsymbol \Psi(z)$, using the Nambu spinors $\boldsymbol \Psi(z) =[{\Psi}_{\uparrow}(z),{\Psi}_{\downarrow}(z),{\Psi}^\dag_{\downarrow}(z),-{\Psi}^\dag_{\uparrow}(z)]^{T}$, where the operator ${\Psi}^\dag_{\sigma}(z)$ creates an electron with spin $\sigma$ at position $z$ in the nanowire. We use
\begin{align}
{\cal H}_{\text{NW}} = \le(-\frac{\hbar^2\p_z^2}{2m^*} - \mu - i\alpha\p_z\sigma_y \ri)\tau_z + \frac{1}{2}g\mu_{\text{B}}{\bf B}_{\rm Z}\cdot \boldsymbol\sigma,
\end{align}
where the Pauli matrices $\boldsymbol\tau$ and $\boldsymbol\sigma$ act in particle-hole space and spin space respectively. Further, $m^*$ is the effective mass of the electrons in the wire, $\mu$ is the chemical potential, $\alpha$ is the Rashba spin-orbit strength, and $g$ is the effective $g$-factor in the wire.

The Hamiltonians for the two superconductors read similarly $H_n = \frac{1}{2}\int d{\bf r}\, \boldsymbol \psi_n^\dag({\bf r}) {\cal H}_{n} \boldsymbol \psi_n({\bf r})$, where the Nambu spinor $\boldsymbol \psi_n({\bf r})$ describes the electrons in superconductor $n$. The BdG Hamiltonian reads
\begin{align}
{\cal H}_{n} = \le(\frac{p_n^2}{2m_{\rm S}^*} - \mu_{\rm S} \ri) \tau_z + \Delta \le[\cos(\phi_n)\tau_x + \sin(\phi_n) \tau_y \ri],
\end{align}
where $p_n$ is the momentum operator for electrons in superconductor $n$, $m_{\rm S}^*$ is their effective mass, and $\mu_{\rm S}$ is the chemical potential of the two superconductors. The superconducting phases are $\phi_1 = -\phi/2$ and $\phi_2 = \phi/2$, corresponding to a phase difference $\phi$.

Finally, the tunnel coupling between the nanowire and the two superconductors is described by the tunneling Hamiltonian
\begin{align}
H_{\text{t}} = \sum_{n,\sigma} \int d{\bf r}\, t_n({\bf r}) \psi^\dagger_{n,\sigma}({\bf r}) \Psi_\sigma(z) + {\rm H.c.},
\end{align}
using the functions $t_1({\bf r}) = t\,\delta(x)\delta(y)\Theta(z)\Theta(L_1-z)$ and $t_2({\bf r}) = t\,\delta(x)\delta(y)\Theta(z-L+L_2)\Theta(L-z)$, with $t$ parameterizing the coupling strength.

Our aim is to calculate the differential conductance of this system, in a setup where one end of the wire is connected to a normal-metal tunnel probe (see Fig.~\ref{fig1}).
We start by writing an expression for the retarded Green function $G^R(z,z',\epsilon)$ for the electrons and holes in the nanowire.
A convenient approach is to first integrate out the degrees of freedom of the two superconductors, resulting in an energy-dependent self-energy for the electrons in the wire. This self-energy (i) describes the proximity effect, i.e.~it introduces (superconducting) correlations between electrons and holes in the wire, and (ii) leads to a renormalization of all energy levels and quasi-particle weights in the wire. Both effects would not be captured fully by a model where the $s$-wave pairing is introduced phenomenologically as a constant pairing potential~\cite{PhysRevB.84.144522}. Recent experiments showed a large and hard induced superconducting gap in epitaxially grown superconductor-semiconductor nanowires~\cite{Chang2015}, and especially in that case retaining an energy-dependent self-energy can make a qualitative and important difference~\cite{PhysRevLett.114.106801}.

We assume that the Fermi energy in the superconductor is by far the largest relevant energy scale in the problem. This allows us to write the self-energy for the electrons and holes in a wire coupled to a bulk superconductor with pairing potential $\Delta e^{i\phi}$ as~\cite{PhysRevB.91.165425}
\begin{equation}
\Sigma_{\rm S} (\Delta,\phi;\epsilon) = \gamma\frac{-\epsilon + \Delta \le[ \cos(\phi)\tau_x + \sin(\phi)\tau_y\ri] }{\sqrt{\Delta^2-(\epsilon+i0^+)^2}},\label{eq:se}
\end{equation}
where $\gamma$ parametrizes the strength of the coupling to the superconductor (it is proportional to $t^2$ and to the normal-state density of states in the superconductor, and corresponds roughly to the normal-state tunneling rate of electrons into the superconductor at the Fermi level) and $0^+$ is a positive infinitesimal.

The self-energy (\ref{eq:se}) is diagonal in position space, and for our setup (involving two superconductors) we thus find the total self-energy
\begin{equation}
\Sigma(z,\epsilon) = 
\begin{cases}
\Sigma_{\rm S} (\Delta,-\phi/2; \epsilon) & \text{for } 0<z<L_1, \\
\Sigma_{\rm S} (\Delta,\phi/2; \epsilon) & \text{for } L-L_2<z<L, \\
0 & \text{otherwise}.
\end{cases} 
\end{equation}
With this, we can write the Green function for the electrons and holes in the wire as
\begin{equation}
G^R(z,z',\epsilon) = \le[\frac{1}{\epsilon-H_{\text{NW}}-\Sigma+i0^+}\ri]_{z,z'}.
\end{equation}

From the Green function, we can derive an expression for the scattering matrix of the nanowire. With only one lead connected, it suffices to consider the reflection matrix of the system,
\begin{align}
R(\epsilon) &= \twomatrix{r_{\text{ee}}(\epsilon)}{r_{\text{eh}}(\epsilon)}{r_{\text{he}}(\epsilon)}{r_{\text{hh}}(\epsilon)} \nonumber\\
&= 1-2\pi i W^\dag \le\{\le[G^R(\epsilon)\ri]^{-1}+i\pi W W^\dag\ri\}^{-1} W.\label{eq:re}
\end{align}
The indices `e' and `h' at the submatrices $r(\epsilon)$ stand for electron and hole, such that, for instance, the matrix $r_{\rm ee}(\epsilon)$ describes normal reflection from the sample and the matrix $r_{\rm eh}(\epsilon)$ describes Andreev (electron-hole) reflection. The second line shows how this reflection matrix can be expressed in terms of the retarded Green function of the electrons in the wire~\cite{Aleiner2002,beenakker:arxiv}. The matrix $W$ contains coupling elements describing the coupling between the basis states in the wire and the modes in the lead.

Finally, in terms of this reflection matrix the leading order (linear) differential conductance at zero temperature can be written as~\cite{PhysRevB.55.3146}
\begin{equation}
\frac{d I}{d V} = \frac{e^2}{h}\text{Tr}\le[1-|r_{\text{ee}}(eV)|^2+|r_{\text{eh}}(eV)|^2\ri].\label{eq:didv}
\end{equation}
The term $(1-|r_{\text{ee}}|^2)$ yields the contribution from normal electron transmission, and $|r_{\text{eh}}|^2$ the contribution from Andreev reflection.

For our numerical calculations, we discretize the Hamiltonian $H_{\rm NW}$ on a one-dimensional lattice with $N$ sites, and the Green function $G^R(\epsilon)$ is then found from inverting the $4N\times 4N$ matrix $[\epsilon - H_{\rm NW} - \Sigma + i0^+]$.
The coupling matrix $W$ is constructed as
\begin{align}
  W = \sqrt{\gamma_W} \le(\mathbf{s}_1\otimes\mathds{1}_4\ri)^{\text{T}},
\end{align}
where $\mathbf{s}_1=(1,0,0,\ldots)$ is a vector of length $N$, $\mathds{1}_4$ is a $4\times 4$ identity matrix and $\gamma_W$ parametrizes the strength of the coupling to the lead. This $4N \times 4$ matrix thus describes a coupling between one single mode in the lead and site $1$ only.
We see from (\ref{eq:re}) that $WW^\dagger$ plays the role of a self-energy for the particles in the wire, so $\gamma_W$ can be seen as the electronic tunneling rate through the barrier between the wire and the lead.
The reflection matrix $R(\epsilon)$ then follows straightforwardly and from that the differential conductance can be calculated using Eq.~(\ref{eq:didv}).

We express all energies in terms of the spin-orbit energy $E_{\text{SO}}\equiv \alpha^2m^*/2\hbar^2 \approx 68\ \mu$eV, where the Rashba spin-orbit strength is set to $\alpha=0.2$ eV\AA{} and the effective mass of the electrons in the nanowire is assumed to be $m^*=0.026m_{\text{e}}$, which is the value for bulk InAs at room temperature.
Unless stated otherwise, the parameters we use in our simulations are as follows: The number of sites in the tight-binding model $N=100$, the length of the wire $L=1.5\ \mu$m, the length of the two superconductors $L_1=L_2=675$ nm, the bulk superconductor gap $\Delta = 2E_{\text{SO}}$, the wire's chemical potential $\mu=0$, the coupling parameter $\gamma =2.5E_{\text{SO}}$, and the coupling of the wire to the lead $\gamma_W=25E_{\text{SO}}$.
This combination of parameters results in an effective hopping matrix element $t=\hbar^2/2m^*a^2 \approx 95 E_{\text{SO}}$, where $a=L/N$ is the lattice constant, and a spin-orbit-induced `spin-flip' nearest-neighbor coupling of $s = \alpha / 2a \approx 9.8E_{\text{SO}}$.

\section{Results\label{sec:results}}

\begin{figure}
\begin{center}
  \includegraphics[scale=1]{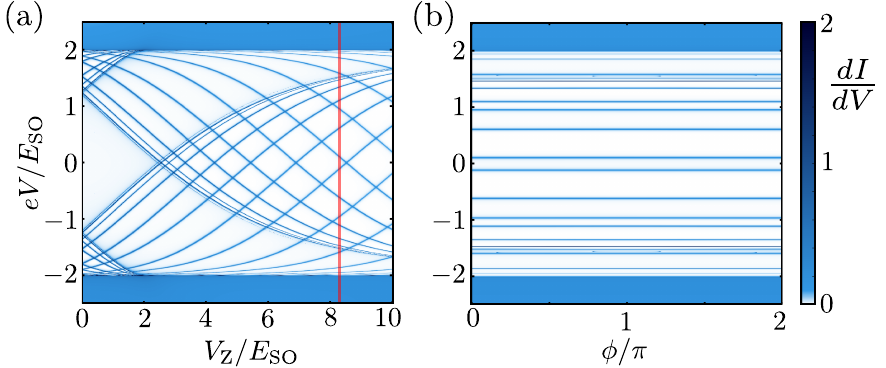}
  \caption{(Color online) Differential conductance of the nanowire in units of $e^2/h$ for $\theta = \pi/2$. With this orientation of the Zeeman field the system is always in the trivial regime. (a) Conductance as a function of bias voltage $V$ and $V_{\rm Z}$ for $\phi=0$. (b) Conductance as a function of $V$ and $\phi$ for $V_{\text{Z}}=8.3E_{\text{SO}}$, indicated by the red line in (a).\label{fig2}}
\end{center}
\end{figure}
We first investigate the differential conductance of the system in the topologically trivial phase, i.e.~where the Zeeman field ${\bf B}_{\rm Z}$ is parallel to the spin-orbit field ($\theta=\pi/2$). In Fig.~\ref{fig2} we plot the differential conductance in units of $e^2/h$ as a function of bias voltage $V$. Fig.~\ref{fig2}(a) shows the dependence on the strength of the Zeeman field $V_{\rm Z} = \tfrac{1}{2} g\mu_{\rm B}|{\bf B}_{\rm Z}|$, assuming no phase difference, $\phi = 0$. We see that conductance is high for energies larger than the bulk superconducting gap, $|eV| > \Delta$. Indeed, at these energies the superconductors have a finite single-particle density of states which allows for tunneling of electrons from the lead into the superconductors. For energies smaller than the bulk gap we see several sharp peaks in the conductance. These peaks are caused by Andreev bound states in the wire which can lead to Andreev reflection of electrons at the lead-wire interface, effectively resulting in the transfer of Cooper pairs from the lead into the superconductors. At $V_{\rm Z} = 0$, there is an induced gap in the wire of $\Delta_{\rm ind} \approx 1.2E_{\rm SO}$, which agrees with the estimate $\Delta_{\rm ind} \sim \Delta\gamma / (\Delta + \gamma)$~\cite{PhysRevB.84.144522}. When $V_{\rm Z}$ increases, the subgap states acquire a Zeeman splitting resulting in a closing of the induced gap at $V_\text{Z}\approx 3E_{\text{SO}}$. The gap does not reopen again, since the nanowire stays in the topologically trivial phase for $\theta = \pi/2$. The bending of the peaks for energies close to $\pm \Delta$ is due to the renormalization of the energy levels in the wire by the proximity of the superconductor. If we would have used a phenomenological (energy-independent) induced gap in the wire Hamiltonian, this bending would have been absent.

We would like to point out that this Zeeman-induced subgap structure has qualitative features in common with the subgap spectrum in the topological regime. After the gap closing, the spacing between the conductance peaks at zero bias, $V = 0$, becomes proportional to $\sqrt{V_{\rm Z}}$, which is the same as the expected $V_{\rm Z}$-dependence of the period of the oscillations due to a finite overlap of the Majorana end states in the topological regime~\cite{PhysRevB.86.220506,Rainis2013}.
Also, the amplitude of the oscillations of the ``splitting of the zero-bias peak'' grows steadily with increasing $V_{\rm Z}$, as it is expected to do in the Majorana case.
Therefore, if one would focus only on the $V_{\rm Z}$-dependence of the lowest-energy peaks in the conductance (for instance, when other features of the subgap structure are smeared by noise or finite temperature), then it could be hard to distinguish the topological from the trivial regime.

We now proceed by investigating the dependence of the conductance on the phase difference $\phi$ between the superconductors. In Fig.~\ref{fig2}(b) we plot the differential conductance as a function of $V$ and $\phi$ with fixed $V_{\rm Z} = 8.3E_{\rm SO}$, which is at the position of the red line in Fig.~\ref{fig2}(a).
We see that the positions of the peaks in the conductance do not depend significantly on $\phi$. Indeed, in the trivial regime the Andreev bound states that are most strongly influenced by the phase difference between the two parts of the wire are expected to live mainly in the central normal region. The tunnel probe is attached to the left end of the wire and is thus predominantly coupled to the Zeeman-split subgap states in the left proximity-induced superconductor. We emphasize the importance of the weak coupling between the tunnel probe at the end of the wire and the Andreev states the middle of the wire for clearly distinguishing the topologically trivial and non-trivial phases in experiment. This requires to have the length of the wire covered by superconductor $S1$ longer than the coherence length of the resulting induced superconductivity. In experiment, one should verify this first by showing that, in the topologically trivial phase, the conductance has virtually no dependence on the phase difference between the superconductors.

\begin{figure}[t]
\begin{center}
  \includegraphics[scale=1]{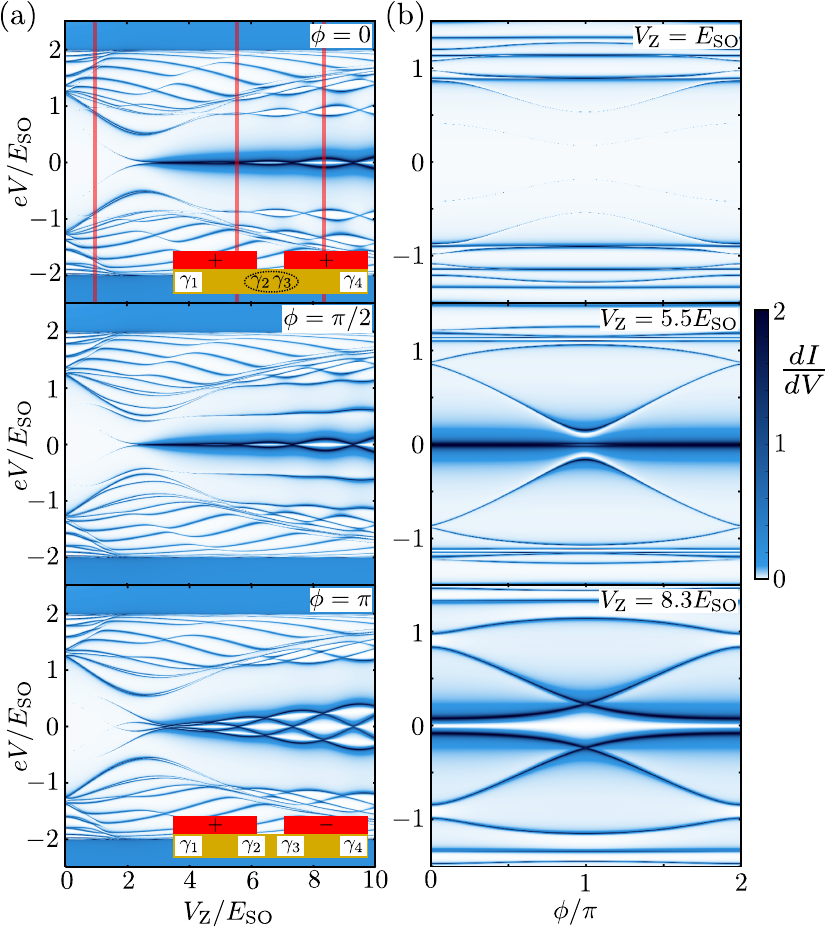}
  \caption{(Color online) Differential conductance of the wire in units of $e^2/h$ for $\theta = 0$. Now the wire can enter a topological phase when increasing $V_{\rm Z}$. (a, left column) Conductance as a function of $V$ and $V_{\rm Z}$ for $\phi = 0$, $\pi/2$, and $\pi$. (b, right column) Conductance as a function of $V$ and $\phi$ for $V_{\rm Z} = E_{\rm SO}$, $5.5 E_{\rm SO}$, and $8.3 E_{\rm SO}$, indicated by the red lines in the top panel of (a).\label{fig3} }
\end{center}
\end{figure}
Next we investigate the conductance of the wire when ${\bf B}_{\rm Z}$ is oriented parallel to the wire, i.e.~$\theta = 0$. In Fig.~\ref{fig3}(a) we plot the differential conductance in units of $e^2/h$ as a function of $V$ and $V_{\rm Z}$ for the phase differences $\phi = 0$, $\pi/2$, and $\pi$ (from top to bottom). We see that after the induced gap closes at $V_{\rm Z} = V_c \approx 2.5 E_{\rm SO}$, it now reopens again, which signals a topological phase transition. For $\phi = 0$ (top plot) and $V_{\rm Z} > V_c$ a strong zero-bias peak appears which at higher $V_{\rm Z}$ splits again and starts to oscillate. This zero-bias peak of conductance $2e^2/h$ as well as its splitting and the oscillations at higher field are consistent with the formation of Majorana end states at the boundaries of the topological regimes in the wire~\cite{PhysRevB.86.220506,Rainis2013}. In our setup, one could in principle expect four Majorana states in the topological phase, as indicated in the cartoon inside the plot. The Majorana states labeled $\gamma_2$ and $\gamma_3$ are however strongly coupled through the central normal region, and thus will gap out to form a normal fermionic state. The two other Majorana modes ($\gamma_1$ and $\gamma_4$) are separated by $L$ and both are localized on the scale of the coherence length $\xi_{\rm M}$. This coherence length is a function of $V_{\rm Z}$ and the overlap of the two Majorana wave functions increases with increasing $V_{\rm Z}$. This explains the splitting of the zero-bias peak into two, and the oscillations are due to the oscillatory form of the Majorana wave functions.
In the case the two superconductors have opposite phases ($\phi = \pi$, bottom plot) the situation is different.
The effective $p$-wave pairing term now changes sign across the central normal region, creating two {\it different} topological phases in the two halves of the wire (this induced $p$-wave superconductor falls in the BDI symmetry class).
Since the Majorana modes $\gamma_2$ and $\gamma_3$ now belong to different phases, they cannot recombine into a normal fermionic mode and gap out, and we thus expect to have {\it four} low-energy Majorana modes in the system~\cite{ramon1,ramon3,Note1}, which will couple and split due to the finite size of the wire.
The pair of Majorana states $\gamma_1$ and $\gamma_2$ (as well as the pair $\gamma_3$ and $\gamma_4$) still belongs to the same topological phase. The states are separated by a distance of roughly $L/2$ and their energies will split when the two wave functions start to overlap significantly.
We indeed see in our simulations two interlaced sets of oscillations which start at lower $V_{\rm Z}$ than those in the top plot and have both a larger period and amplitude.
All these differences can be explained by the reduced separation of the overlapping Majorana states.
The middle plot ($\phi = \pi/2$) shows an intermediate situation, where the two central modes $\gamma_2$ and $\gamma_3$ are still coupled, but the coupling is not strong enough to push their energies outside of the induced gap.

In Fig.~\ref{fig3}(b) we plot the differential conductance as a function of $V$ and the phase difference $\phi$, for three different Zeeman fields: $V_{\rm Z} = E_{\rm SO}$ (before the closing of the gap) and $V_{\rm Z} = 5.5E_{\rm SO}$ and $8.3E_{\rm SO}$ (after the reopening of the gap), indicated by the red lines in the top left plot. In contrast to the topologically trivial case shown in Fig.~\ref{fig2}(b), the low-energy conductance peaks now show a distinctive dependence on the phase difference. In the lower two plots we see how the peak due to the lowest excited state moves towards zero energy and increases in intensity when the phase difference goes from $0$ to $\pi$. This is the gapped fermionic mode formed by the two central Majorana states gradually developing into two uncoupled low-energy Majorana modes with significant weight at the ends of the wire, cf.~the spectra of Andreev bound states presented in Refs.~\cite{ramon1,ramon2,ramon3}.

\begin{figure}
\begin{center}
  \includegraphics[scale=1]{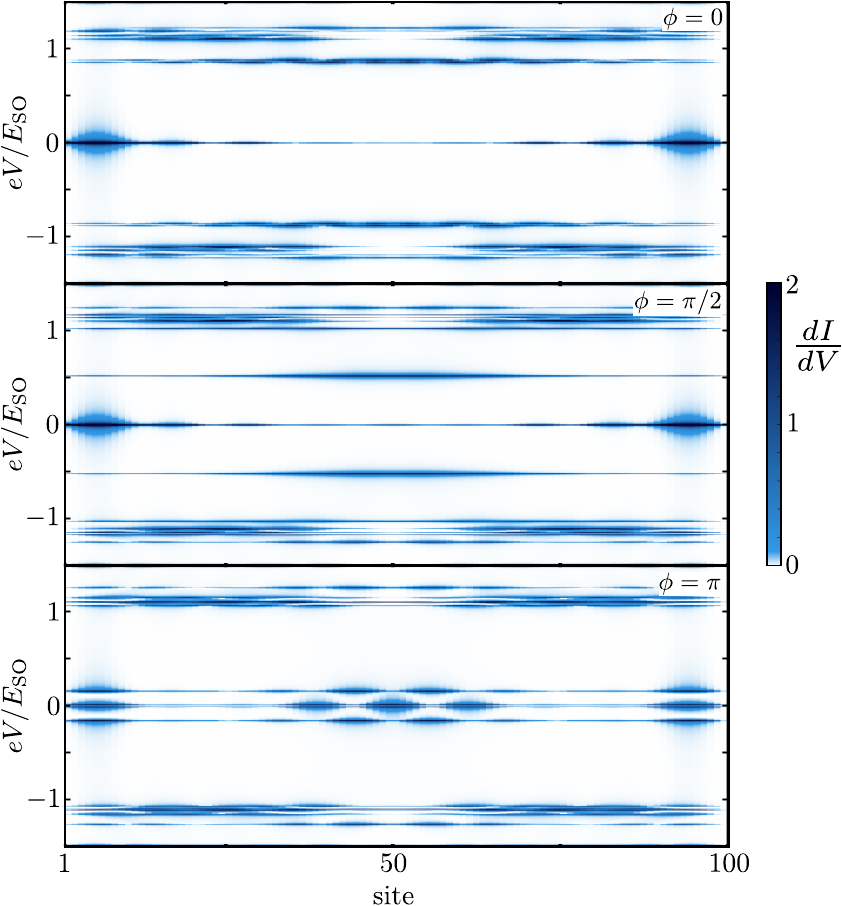}
  \caption{(Color online) The differential conductivity of the wire in units of $e^2/h$ as a function of $V$ and position of the normal probe lead. We have set the phase difference to $\phi = 0$, $\pi/2$, and $\pi$ (from top to bottom), the Zeeman energy to $V_{\text{Z}}=5.5E_{\text{SO}}$, and the coupling strength to the lead to $\gamma_W=E_{\text{SO}}$.\label{fig4}}
\end{center}
\end{figure}
We can support this picture by plotting the local differential conductivity (where we vary the position of the probe lead along the wire, encoded in the matrix $W$), which maps out the local (tunneling) density of states in the wire. We show the result in Fig.~\ref{fig4} for $V_{\text{Z}}=5.5E_{\text{SO}}$ and three different phase differences, $\phi = 0$, $\pi/2$, and $\pi$. The lowest excited state indeed develops from a bulk state with energy $\sim E_{\rm SO}$ at $\phi = 0$ to a low-energy state localized at the boundaries of the topological regimes at $\phi = \pi$. This behavior is generally seen for all values of Zeeman fields in the topologically non-trivial phase.

The plots presented in Fig.~\ref{fig3} are the main results of our numerical work. They illustrate how the low-energy Majorana modes should manifest themselves as strong peaks in the differential conductance with a distinctive dependence on the phase difference $\phi$, which is absent in the trivial case, see Fig.~\ref{fig2}(b). The main feature that can be discerned is the gradual gapping out of the two central Majorana modes when $\phi$ is changed from $\pi$ to $0$. Below, in Sec.~\ref{sec:effective_model}, we will investigate the phase-dependence of the low-energy part of the spectrum in more detail and present a simple model to analytically understand the level structure as a function of $\phi$.

\begin{figure}
  \includegraphics[width=\columnwidth]{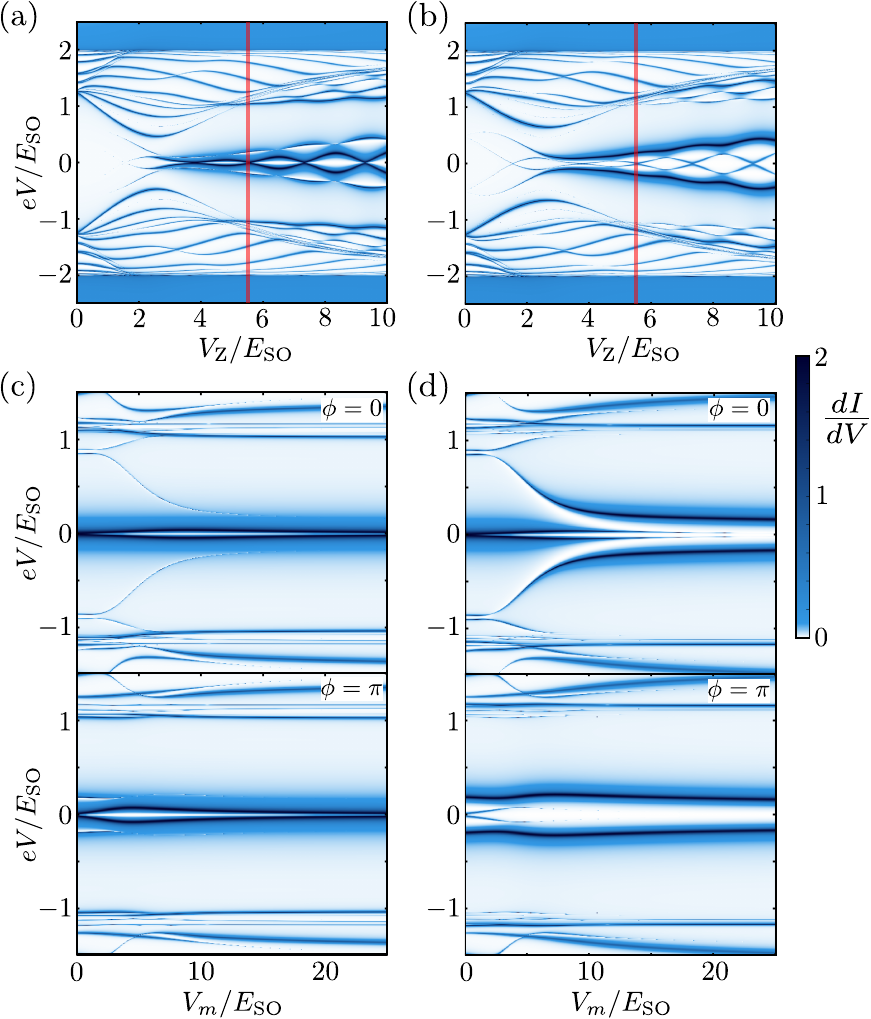}
  \caption{(Color online) Differential conductance in units of $e^2/h$ for the case that $L_1\neq L_2$. We took $\theta=0$ and $\phi=\pi$, and show $dI/dV$ as a function of $V$ and $V_{\rm Z}$ (a,b) and of $V$ and the potential of the central normal region $V_m$ (c,d). In (a,c) we have set $L_1=825$~nm and $L_2=525$~nm and in (b,d) we took the opposite, $L_1=525$~nm and $L_2=825$~nm. In (a,b) we did not include an extra potential offset in the central region, $V_m = 0$, and in (c,d) we fixed $V_{\rm Z} = 5.5E_{\rm SO}$, as indicated by the red lines in the top plots.\label{fig5}}
\end{figure}
However, before we move on to present our low-energy model, we first investigate how small deviations from the idealized system pictured in Fig.~\ref{fig1} would affect the conductance spectrum. Firstly, in an experimental setup it is unlikely that the two superconductors will be of exactly the same length. In Fig.~\ref{fig5}(a,b) we show the differential conductance as a function of $V$ and $V_{\rm Z}$ for $L_1 > L_2$ and $L_1 < L_2$ respectively. We see that when the two superconductors are of different lengths, the general structure can still be similar to that of Fig.~\ref{fig3}(a, bottom plot), but now anti-crossings arise between some of the low-energy modes. In Fig.~\ref{fig5}(a) where $L_1>L_2$ we see that the ``inner'' modes show a higher conductance than the ``outer'' ones, and we see the opposite in Fig.~\ref{fig5}(b) where $L_1<L_2$. Indeed, since the tunneling lead is connected to the left end of the wire, where it probes the local density of states, this signifies that the inner modes are localized mainly in the longer superconducting part of the wire and the outer modes mainly in the shorter part.

Another ingredient in an experimental setup could be a tunable potential barrier in the central normal region. We add this complication to our model as a constant potential of height $V_m$ on all lattice sites where $\Sigma = 0$. Raising this barrier can deplete the wire in the normal region and gradually reduce the overlap of the two central Majorana modes. As the barrier becomes high enough this effectively cuts the wire into two uncoupled sections. In Fig.~\ref{fig5}(c,d) we show the conductance spectrum as a function of $V_m$, again for $L_1 > L_2$ and $L_1 < L_2$ respectively, and for two phase differences $\phi = 0$ and $\pi$. We see that the two central Majorana modes that are gapped out at $V_m = 0$ and $\phi = 0$ (top plots) indeed move towards zero energy when the barrier is increased, due to their suppressed overlap. They hybridize with the original low-energy Majorana mode localized at the end of their section of the wire. At the right end of the plots in Fig.~\ref{fig5}(c,d) the wire is effectively cut in two by the high barrier in the middle and the splitting of the two Majorana modes in the left section depends on both the length of the section and the Zeeman field (the shorter section will in general show a larger splitting due to a larger overlap between the MBSs). At $\phi = \pi$ the two central modes are always uncoupled and the dependence of the conductance spectrum on $V_m$ is less pronounced (bottom plots).  Since $V_m$ could be easily tuned by adding an extra gate electrode to the sample, investigating the barrier- and phase-dependent conductance spectrum (such as in Fig.~\ref{fig5}) could be used as an alternative direction in the search for signatures of Majorana physics in the wire.

\section{Low-energy model\label{sec:effective_model}}

As presented in Sec.~\ref{sec:results}, our numerical calculations produced low-energy peaks in the differential conductance of a topological N-SNS junction and we showed that these peaks can be associated with states living on the boundaries between topologically trivial and non-trivial regions, all consistent with the interpretation of these states as being Majorana bound states. In the present Section we will provide further support for this picture by investigating a one-dimensional spinless $p$-wave superconductor with a phase discontinuity. We develop an effective model to describe the low-energy physics of this system and we show how it produces up to four Majorana bound states (at the ends of the system and at the phase discontinuity). We then map the parameters in the $p$-wave Hamiltonian to those in the nanowire Hamiltonian (\ref{eq:hw}) and show how this simple toy model produces qualitatively the same phase-dependent features as those found from our numerical simulations in Sec.~\ref{sec:results}.

\begin{figure}
  \includegraphics[width=\columnwidth]{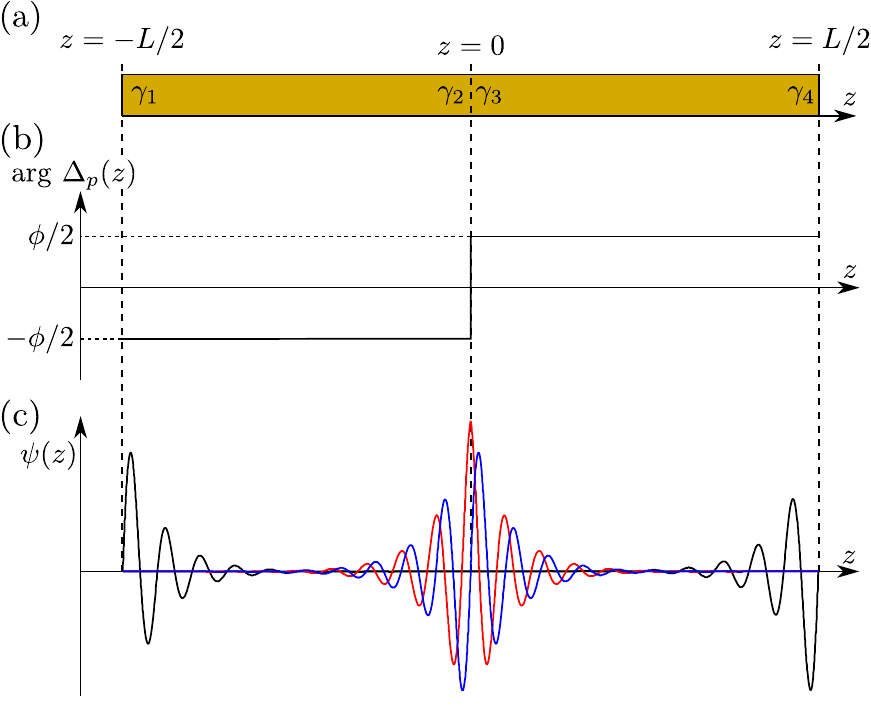}
  \caption{(Color online) (a) Sketch of a one-dimensional $p$-wave superconductor centered around $z=0$. We expect Majorana bound states to form at the ends of the wire and close to the phase jump at $z=0$. (b) The phase of the pairing potential $\Delta_p(z)$ as a function of $z$. (c) \label{fig6}}
\end{figure}

We thus consider a one-dimensional $p$-wave superconductor which we describe with the Hamiltonian $H_p = \int dz\,\boldsymbol \psi^\dagger(z){\cal H}_p\boldsymbol \psi(z)$, where we use Nambu spinors in particle-hole space $\boldsymbol \psi(z) = [ \psi(z), \psi^\dagger(z) ]^T$. For ${\cal H}_p$ we use the simple model Hamiltonian~\cite{bernevig-book}
\begin{equation}
  {\cal H}_p = 
  \left(\begin{array}{cc}
         -\frac{\hbar^2\p_z^2}{2m}-\mu & -\tfrac{i}{2}\{ \Delta_p(z),\p_z\} \\
         -\tfrac{i}{2}\{ \Delta^*_p(z), \p_z \} & \frac{\hbar^2\p_z^2}{2m}+\mu
        \end{array}
  \right),\label{eq:hpw}
\end{equation}
where the pairing potential is position-dependent,
\begin{equation}
\Delta_p(z) =
\begin{cases}
\Delta_p e^{-i\phi/2} & \text{for } z<0, \\
\Delta_p & \text{for } z=0, \\
\Delta_p e^{i\phi/2} & \text{for } z>0.
\end{cases} 
\end{equation}
The superconductor is assumed to be of length $L$ and the center of the wire corresponds to $z=0$, see Fig.~\ref{fig6}(a). As sketched in Fig.~\ref{fig6}(b), the phase of the superconducting pairing potential $\Delta_p(z)$ jumps from $-\phi/2$ to $\phi/2$ at $z=0$. This $p$-wave superconductor is in the topologically non-trivial phase for $\mu>0$ \cite{bernevig-book}, which is the only case we will consider in this Section.

We expect this model to describe a situation similar to the one investigated in Sec.~\ref{sec:results}: For $\phi=0$, the wire should have two Majorana bound states, one at each end, with an exponentially small energy splitting $\delta\epsilon \propto e^{-L/l_c}$~\cite{PhysRevB.86.220506}, where $l_c$ is the coherence length of the superconductor. For $\phi \to \pi$ we expect two additional Majorana states to form close to the phase discontinuity at $z=0$. We will now analyze the $p$-wave Hamiltonian (\ref{eq:hpw}) and try to derive an effective low-energy model to describe the physics of the four Majorana levels.

We try to solve the Schr{\"o}dinger equation
\begin{equation}
({\cal H}_p-E)\psi(z)=0, \label{eq:sche}
\end{equation}
using for $\psi(z)$ the Ansatz
\begin{equation}
\psi(z) =\sum_n e^{c_n z}\twovector{u_n}{v_n}.
\end{equation}
This yields four solutions for (\ref{eq:sche}) with
\begin{equation}
  c_n = \pm \frac{1}{\xi}
  \sqrt{\tfrac{1}{2}-\tilde{\mu}\pm i \sqrt{\tilde{\mu}-\tilde{E}^2-\tfrac{1}{4}}},\label{eq:ref1}
\end{equation}
where we use the coherence length $\xi = \hbar^2 / 2m\Delta_p$ and introduced the dimensionless $\tilde E = \xi E / \Delta_p$ and $\tilde \mu = \xi \mu / \Delta_p$, and we assumed that $\tilde \mu > \tilde E^2 +\tfrac{1}{4}$. In fact, we will take $\tilde \mu \gg 1$ which is deep in the topological regime and allows for several convenient simplifications. With (\ref{eq:ref1}) we find
\begin{equation}
v_n^\pm = i\frac{\tilde c_n^2 + \tilde E + \tilde \mu}{\tilde c_n e^{\pm i \phi/2}}u_n^\pm,\label{eq:vn}
\end{equation}
where the $+$ applies to the right part of the wire, where $z>0$, and the $-$ to the left part, $z<0$. We also used a normalized $\tilde c_n = \xi c_n$.

We expect the bound states to be localized close to $z = \pm L/2$ and $z=0$ (depending on $\phi$). In principle one should thus allow all four solutions for $c_n$ in both the left and right part of the wire, and then find the explicit wave functions and energies from imposing the right boundary conditions and matching the solutions at $z=0$.

Since we cannot solve this problem analytically, we construct approximate solutions by separating the problem into two parts. First we focus only on the bound states around $z=0$, and we thus assume the wire to be infinitely long. Now we can only have decaying solutions for increasing $|z|$, which means that for $z>0$ only the two $c_n$ with ${\rm Re}\,c_n <0$ are allowed and for $z<0$ only the two with ${\rm Re}\,c_n >0$. We then proceed as follows: (i) We match the wave functions at $z=0$. (ii) Since the derivative can be discontinuous due to the $\delta$-function resulting from the term $\partial_z \Delta_p(z)$ in the Hamiltonian, we impose
\begin{equation}
\lim_{\eta\to 0} \int_{-\eta}^\eta dz\, ({\cal H}_p - E)\psi(z) = 0.
\end{equation}
(iii) We require normalized solutions, $\int dz\, |\psi(z)|^2 = 1$. This all together allows us to find explicit solutions for the eigenstates and -energies for the two lowest levels. We make use of the assumption $\tilde\mu \gg 1$, and after a $\pi/2$-rotation along $\tau_x$ in particle-hole space we can finally write the Hamiltonian for this $2\times 2$ subspace as
\begin{equation}
H_{23} = \frac{\Delta_p\sqrt{\tilde \mu}}{\xi}\cos(\tfrac{\phi}{2})\tau_y,\label{eq:h23}
\end{equation}
and find the wave functions
\begin{align}
 \psi_2 & = \sqrt{\frac{|\sin\tfrac{\phi}{2}|}{2}}
 e^{-\tfrac{1}{2\xi}|z\sin\frac{\phi}{2}|}
 \cos\left(\tfrac{\sqrt{\tilde \mu}}{\xi}z\right)
 \left(\begin{array}{c} i \\ -i \end{array}\right), \\
 \psi_3 & = \sqrt{\frac{|\sin\tfrac{\phi}{2}|}{2}}
 e^{-\tfrac{1}{2\xi}|z\sin\frac{\phi}{2}|}
 \sin\left(\tfrac{\sqrt{\tilde \mu}}{\xi}z\right)
 \left(\begin{array}{c} i \\ -i \end{array}\right),
\end{align}
cf.~Ref.~\cite{kwon}.
These two wave functions oscillate with period $\xi/\sqrt{\tilde \mu}$ and decay exponentially on a length scale $2\xi/|\sin\frac{\phi}{2}|$. The prefactor $(\frac{1}{2}|\sin\frac{\phi}{2}|)^{1/2}$ follows from normalization of the wave functions. We see that for $\phi = \pi$ the two states are strongly localized around $z=0$ and their splitting is zero, whereas for $\phi \to 0$ the localization length diverges, their splitting becomes maximal, and the wave functions look more like bulk modes. This all is in agreement with the picture we presented in Sec.~\ref{sec:model}. In Fig.~\ref{fig6}(c) we sketch the $z$-dependence of the two wave functions for $\phi = \pi$ and $\tilde \mu = 20$, showing the prefactor of the spinor $[i,-i]^T$ (blue and red curves in the central part of the plot).

We now turn to the bound states at the ends of the wire, at $z=\pm L/2$. As an approximation, we ignore the phase jump at $z=0$: Terms in the wave function that decay for increasing $z$ live almost entirely in the left part of the wire, and we assume that for these terms Eq.~(\ref{eq:vn}) applies with a minus sign for the whole extent of the wire, i.e.\ these terms are constructed taking $\Delta_p(z) = \Delta_p e^{-i\phi/2}$ throughout the wire. Similarly, we set $\Delta_p(z) = \Delta_p e^{i\phi/2}$ everywhere for all terms decaying with decreasing $z$. Using the boundary conditions $\psi(\pm L/2) = 0$ and the normalization constraint $\int dz\, |\psi(z)|^2 = 1$ we can again arrive at analytic expressions for the eigenstates and -energies. In the limit of $\tilde\mu \gg 1$ and after an appropriate rotation in particle-hole space, we find for the two lowest levels the Hamiltonian
\begin{equation}
H_{14} =
-2\frac{\Delta_p\sqrt{\tilde \mu}}{\xi}
\cos(\tfrac{\phi}{2})
\sin\left(\tfrac{\sqrt{\tilde \mu}}{\xi}L\right)
e^{-\tfrac{L}{2\xi}}\tau_y,\label{eq:h14}
\end{equation}
and the wave functions
\begin{align}
 \psi_1 & = \frac{e^{-\tfrac{1}{2\xi}(z+L/2)}}{\sqrt 2}
 \sin\left(\tfrac{\sqrt{\tilde \mu}}{\xi}[z+\tfrac{L}{2}]\right)
 \left(\begin{array}{c} e^{i\pi/4-i\phi/4} \\ e^{-i\pi/4+i\phi/4} \end{array}\right), \\
 \psi_4 & = \frac{e^{\tfrac{1}{2\xi}(z-L/2)}}{\sqrt 2}
 \sin\left(\tfrac{\sqrt{\tilde \mu}}{\xi}[z-\tfrac{L}{2}]\right)
 \left(\begin{array}{c} e^{-i\pi/4+i\phi/4} \\ e^{i\pi/4-i\phi/4} \end{array}\right).
\end{align}
These two wave functions oscillate with the same period $\xi/\sqrt{\tilde \mu}$ as $\psi_{2,3}$ found above, and always decay exponentially on a length scale $\xi$. The splitting between the states vanishes for $\phi = \pi$ and is maximal for $\phi = 0$, similar to the splitting between the central two states. However, since $\psi_{1,4}$ are always localized at the ends of the wire and never acquire a bulk character, their maximal splitting is reduced with a factor $e^{-L/2\xi}$. The black curves localized at $z=\pm L/2$ in Fig.~\ref{fig6}(c) show the typical $z$-dependence of these solutions.

To complete our effective low-energy model, we have to include the coupling between the end states $\psi_{1,4}$ and the central states $\psi_{2,3}$. To that end we consider the two halves of the wire separately assuming that the main contribution to this overlap comes from the regions close to $z = \pm L/4$. The approximate wave functions derived above cannot be used to calculate the matrix elements directly (all leading-order terms in $\tilde\mu$ cancel), and we have to infer the splitting between the end states and the central states from the similarity of their wave functions.

We focus here on the matrix element between $\psi_1$ and $\psi_3$, all other elements follow from similar reasoning. We notice that $\psi_3$ and $\psi_4$ have a very similar structure, the differences being: (i) a renormalized $\xi$ in the exponent in $\psi_3$, (ii) a different, $\phi$-dependent prefactor in $\psi_3$, and (iii) a different orientation in particle-hole space. If we can understand the resulting differences in the matrix element between the two states, we can thus modify $H_{14}$ to describe the coupling between $\psi_1$ and $\psi_3$. The spinor structure of $\psi_1$ and $\psi_3$ together with the structure of the Hamiltonian guarantees that the matrix element has to be proportional to $(\cos\frac{\phi}{4} + \sin\frac{\phi}{4})$. This factor replaces the factor $\cos\frac{\phi}{2}$ in (\ref{eq:h14}), which results in the same way from the spinor structure of $\psi_1$ and $\psi_4$. We assume that the renormalization of $\xi$ in $\psi_3$ mainly results in a different exponential suppression: $e^{-L/2\xi} \to e^{-L(1+|\sin\frac{\phi}{2}|)/8\xi}$, where we used that the length of the segment we consider is now $L/2$. Including the extra prefactor $|\sin\frac{\phi}{2}|^{1/2}$, we thus infer the $2\times 2$ coupling Hamiltonian
\begin{widetext}
\begin{align}
H_{13} = & -2\frac{\Delta_p\sqrt{\tilde \mu}}{\xi}
(\cos\frac{\phi}{4} + \sin\frac{\phi}{4}) \sqrt{|\sin \tfrac{\phi}{2} |}
\sin\left(\tfrac{\sqrt{\tilde \mu}}{2\xi}L\right)
e^{-\tfrac{L}{8\xi}(1+|\sin\frac{\phi}{2}|)}\tau_y.
\end{align}

Exactly the same reasoning yields explicit expressions for the remaining seven matrix elements, and we finally arrive at the approximate low-energy Hamiltonian (cf.~the low-energy Hamiltonian inferred from numerical calculations in the supplementary material in \cite{chiu_arx})
\begin{equation}
H_{\rm M} = 2\frac{\Delta_p\sqrt{\tilde\mu}}{\xi}\left( \begin{array}{cccc}
0 & if_c & if_s & i\cos(\tfrac{\phi}{2}) \sin\left(\tfrac{\sqrt{\tilde \mu}}{\xi}L\right)e^{-L/2\xi} \\
-if_c & 0 & -\tfrac{i}{2}\cos(\tfrac{\phi}{2}) & if_c \\
-if_s & \tfrac{i}{2}\cos(\tfrac{\phi}{2}) & 0 & -if_s \\
-i\cos(\tfrac{\phi}{2}) \sin\left(\tfrac{\sqrt{\tilde \mu}}{\xi}L\right)e^{-L/2\xi} & -if_c & if_s & 0
\end{array}\right),\label{eq:4mm}
\end{equation}
where we used the functions
\begin{align}
f_c = {} & {} (\cos\tfrac{\phi}{4} + \sin\tfrac{\phi}{4}) \sqrt{|\sin \tfrac{\phi}{2} |}
\cos\left(\tfrac{\sqrt{\tilde \mu}}{2\xi}L\right)
e^{-\tfrac{L}{8\xi}(1+|\sin\frac{\phi}{2}|)},\\
f_s = {} & {} (\cos\tfrac{\phi}{4} + \sin\tfrac{\phi}{4}) \sqrt{|\sin \tfrac{\phi}{2} |}
\sin\left(\tfrac{\sqrt{\tilde \mu}}{2\xi}L\right)
e^{-\tfrac{L}{8\xi}(1+|\sin\frac{\phi}{2}|)}.
\end{align}

We would now like to connect this effective low-energy model for the $p$-wave Hamiltonian (\ref{eq:hpw}) to the results presented in Sec.~\ref{sec:results}, i.e.~we would like to express the parameters $\sqrt{\tilde \mu}$ and $\xi$ in terms of the parameters of the nanowire Hamiltonian (\ref{eq:hw}). Using the result for $\phi = 0$ and either small $\alpha$ or $\Delta_{\rm ind}$~\cite{PhysRevB.86.220506}, we can identify
\begin{align}
\frac{\sqrt{\tilde \mu}}{\xi} \equiv {} & {} k^*_{\rm F} = \frac{\sqrt{2m^*}}{\hbar} \sqrt{ \mu + \frac{m^*\alpha^2}{\hbar^2}+ \sqrt{\left(\mu + \frac{m^*\alpha^2}{\hbar^2}\right)^2 + V_{\rm Z}^2 - \Delta_{\rm ind}^2 - \mu^2}},\\
\xi^{-1} = {} & {} \frac{m^*}{\hbar^2} \frac{\alpha\Delta_{\rm ind}}{\sqrt{\left(\mu + \frac{m^*\alpha^2}{\hbar^2}\right)^2 + V_{\rm Z}^2 - \Delta_{\rm ind}^2 - \mu^2}},
\end{align}
\end{widetext}
where $\Delta_{\rm ind}$ is the induced pairing potential in the wire, in this Section for simplicity assumed to be constant. This yields for the energy scale $2\Delta_p \sqrt{\tilde \mu}/\xi = \hbar^2 k^*_{\rm F} / m\xi$, which is consistent with the results of \cite{PhysRevB.86.220506}.

\begin{figure}[t]
  \includegraphics[scale=1]{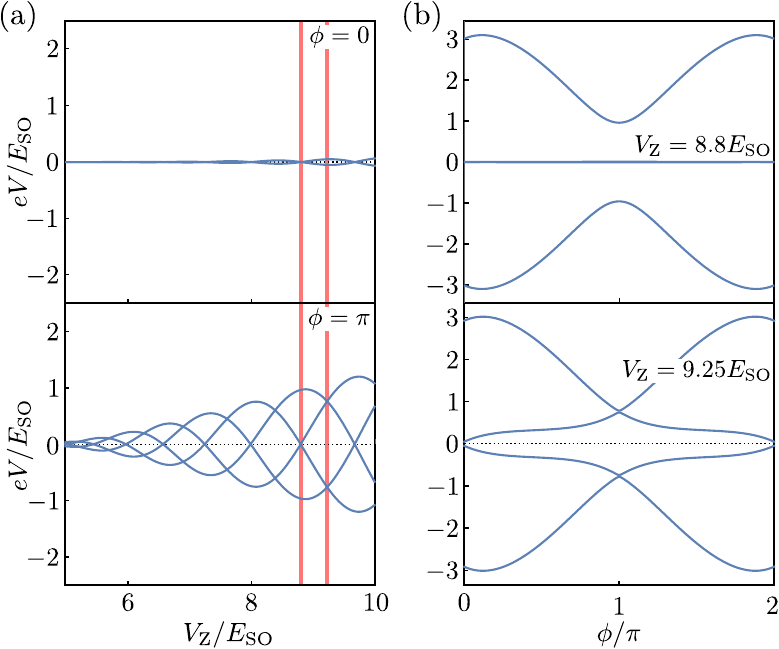}
  \caption{(Color online) The spectrum of $H_{\rm M}$ for $\mu = 0$, $\Delta_{\rm ind} = 4 E_{\rm SO}$, and $L = 20 l_{\rm SO}$. (a) Level structure as a function of $V_{\rm Z}$ for $\phi=0$ (top) and $\phi=\pi$ (bottom). (b) Spectrum as a function of $\phi$ for $V_{\rm Z} = 8.8E_{\rm SO}$ (top) and $V_{\rm Z} = 9.25 E_{\rm SO}$ (bottom), as indicated by the red lines in (a).\label{fig7}}
\end{figure}

In Fig.~\ref{fig7} we plot the spectrum of $H_{\rm M}$, using $\mu = 0$ and $\Delta_{\rm ind} = 4 E_{\rm SO}$ and $L = 20 l_{\rm SO}$ where we define the spin-orbit length as $l_{\rm SO} = \hbar^2/m\alpha$.
Fig.~\ref{fig7}(a) shows the energy levels as a function of $V_{\rm Z}$ for $\phi = 0$ (top plot) and $\phi = \pi$ (bottom plot), to be compared with the top and bottom plots of Fig.~\ref{fig3}(a).
We see that for $\phi = 0$ the two central states are again gapped out, their splitting being $\sim \hbar^2 k^*_{\rm F}/m\xi$.
The splitting of the two remaining low-energy states is suppressed by a factor $e^{-L/2\xi}$ and shows oscillations with a period of $k_{\rm F}^*$.
At $\phi = \pi$ the coupling between $\psi_2$ and $\psi_3$ vanishes, and all energies are suppressed by a factor $e^{-L/4\xi}$.
In Fig.~\ref{fig7}(b) we plot the spectrum as a function of $\phi$, for $V_{\rm Z} = 8.8E_{\rm SO}$ (top plot) and $V_{\rm Z} = 9.25 E_{\rm SO}$ (bottom plot), to be compared with the two lower two plots of Fig.~\ref{fig3}(b).
We see the same distinctive phase-dependence as in our numerical results: When the phase difference goes from $0$ to $\pi$, the gapped fermionic mode formed by the two central Majorana states $\psi_2$ and $\psi_3$ gradually develops into two low-energy Majorana modes that are weakly coupled to $\psi_1$ and $\psi_4$.
The slight bending of the levels close to $\phi = 0$ and $\phi = 2\pi$, which is absent in the numerical results presented in Fig.~\ref{fig3}, is a consequence of including only four levels in the low-energy description:
In the ``full'' numerical model of Sec.~\ref{sec:results}, two of the four low-energy states in fact merge with the above-gap states when $\phi$ approaches a multiple of $2\pi$, and therefore we should not expect to find their correct energies with a low-energy model that does not include these above-gap states.

Finally, we only mention that it is straightforward to modify the Hamiltonian (\ref{eq:4mm}) to describe the case where the lengths of the two parts of the wire are different. The resulting spectrum (not shown here) indeed resembles the low-energy part of Fig.~\ref{fig5}(a,b), reproducing the anti-crossings observed in the figure.

\vspace{0.5cm}
\section{Conclusions\label{sec:conc}}
In conclusion, we investigated the phase-dependent conductance spectrum of a topological N-SNS junction in a semiconducting nanowire. Creating such a system is feasible with the current state of experimental techniques, and we showed that it should in principle allow to determine whether or not a measured low-bias peak in the differential conductance can be associated with non-local fermionic states formed by MBSs at the topological phase boundaries inside the wire. We presented numerical calculations of the conductance spectrum and indicated the phase-dependent features that could serve as a distinguishing signature of the formation of MBSs at the ends of the wire. We supported our interpretation with a simple low-energy model based on a one-dimensional spinless $p$-wave superconductor with a phase discontinuity. This toy Hamiltonian reproduced all important qualitative features of the numerical results and provides insight in the structure of the ``Majorana subspace'' including the gradual gapping out of the central two Majorana states when the phase difference is reduced to zero.

\section{Acknowledgments}

We gratefully acknowledge very helpful discussions with M.~Leijnse, C.~M.\ Marcus, M.~Deng, and S.~M.\ Albrecht. This work was supported by the Sapere Aude program of The Danish Council for Independent Research (EBH).
The Center for Quantum Devices is funded by the Danish National Research Foundation. The research was supported by The Danish Council for Independent Research~\textbar~Natural Sciences.


\end{document}